\documentclass[journal]{IEEEtran} 
\IEEEoverridecommandlockouts
\usepackage{amsmath,amssymb,amsfonts}
\usepackage{algorithmic}
\usepackage{graphicx}
\usepackage{stfloats}
\usepackage{textcomp}
\usepackage{threeparttable}
\usepackage{wasysym}
\usepackage{booktabs}
\usepackage{cite}
\usepackage{comment}
\usepackage{xcolor}
\usepackage{array}
\def\BibTeX{{\rm B\kern-.05em{\sc i\kern-.025em b}\kern-.08em
    T\kern-.1667em\lower.7ex\hbox{E}\kern-.125emX}}

\newlength{\shiftwidth}
\begin{document}

\title{Modular Simulation Environment Towards OTN AI-based Solutions}

\author{
    \IEEEauthorblockN{Sam Aleyadeh\IEEEauthorrefmark{1}, Abbas Javadtalab\IEEEauthorrefmark{2}, and Abdallah Shami\IEEEauthorrefmark{1}} \\
\IEEEauthorblockA{\IEEEauthorrefmark{1}Western University, London, Ontario, Canada; e-mails: \{saleyade,abdallah.shami\}@uwo.ca\\
}
\IEEEauthorblockA{\IEEEauthorrefmark{2}Concordia University, Montreal, Canada; e-mails:\{Abbas.javadtalab\}@concordia.ca}	
}

\maketitle

\begin{abstract}
The current trend for highly dynamic and virtualized networking infrastructure made automated networking a critical requirement. Multiple solutions have been proposed to address this, including the most sought-after machine learning ML-based solutions. However, the main hurdle when developing Next Generation Network is the availability of large datasets, especially in 5G and beyond and Optical Transport Networking (OTN) traffic. This need led researchers to look for viable simulation environments to generate the necessary volume with highly configurable real-life scenarios, which can be costly in setup and require subscription-based products and even the purchase of dedicated hardware, depending on the supplier. We aim to address this issue by generating high-volume and fidelity datasets by proposing a modular solution to adapt to the user’s available resources. These datasets can be used to develop better-aforementioned ML solutions resulting in higher accuracy and adaptation to real-life networking traffic.
\end{abstract}

\begin{IEEEkeywords}
Simulation, Machine learning, 5G, OTN, DataSets
\end{IEEEkeywords}

\section{Introduction \& Motivation}\label{ch4_intro}
\subsection{Motivation}
\indent\indent The adaptation of virtualized networking architecture is currently on the rise due to multiple attractive networking and security factors, such as lower costs, high adaptability, increased robustness, reduction in latency \cite{callegati2014performance}, better intrusion detection, and increased networking anonymity \cite{melkov2021security}. This push shifts the availability and placement of the networking architecture from monitored locations with technicians on hand to a more sparsely located deployment that is meant for a more automated operation approach \cite{akyildiz20206g}. To maintain the robustness in such setups, redundancies and other more advanced techniques are employed. Such approaches included optimization-based techniques focused on network planning  \cite{hawilo2019network}. However, the highly complex nature of the optimization techniques used has limited their usability in real-life deployments. Heuristic-based solutions, such as \cite{aleyadeh2022optimal}, were effective to a certain extent but limited in adapting to changes in the user base or the available networking infrastructure. Ultimately the above solutions have reached their eventual bottleneck, given the increased complexity of the networking architecture and the highly variable nature of the deployed environments \cite{celik2022top}. Artificial Intelligence (AI) based solutions were practical approaches in other venues due to their ability to adapt each instance based on the local deployment environment and continuous dynamic changes to maintain a high efficacy \cite{rafique2018machine}. However, unlike many of their counterparts, networking-based ML solutions lack publicly available datasets. This lack can be traced back to several compounding factors ranging from network operators’ fears for their users’ privacy, their protection of Intellectual property and patented networking architecture, and exploitation of the datasets in finding methods to attack their internal networks \cite{sebbar2021secure}. This scarcity spurred a number of research efforts to create simulation environments to address this challenge, such as \cite{sharkh2016building}; however, such works still could not satisfy the apparent lack of datasets, especially in contemporary networking architectures such as 5G and optical transport networks (OTNs), where new datasets are challenging to find in a volume adequate to perform proper ML testing and debugging. 

\subsection{Available Simulators}
\indent\indent Several recent vendor-based solutions have allowed researchers and companies to progress their work without real-life data \cite{tetcos_2022,oliveira2021generating,networks_2022,sumo}. Their solutions are primarily subscriptions based and may entail a costly initial investment in the form of dedicated hardware, add-on libraries, or features based on the accuracy type and volume of data required. Table \ref{comparisontable} outlines a few of the more popular available solutions along with their features and requirements.

\begin{table*}[ht]
	\centering
	\footnotesize
	\newcommand*\rot[1]{\hbox to1em{\hss\rotatebox[origin=br]{-50}{#1}}}
	\newcommand*\feature[1]{\ifcase#1 -\or\LEFTcircle\or\RIGHTcircle\or\CIRCLE\or\space\or W \or U \or B\fi}
	\newcommand*\f[3]{\feature#1&\feature#2&\feature#3}
	\makeatletter
	\newcommand*\ex[9]{#1\tnote{#2}&#3&%
		\f#4&\f#5&\f#6&\f#7&\f#8&\expandafter\f\@firstofone
	}
	\makeatother
	\newcolumntype{G}{c@{}c@{}c}
	\begin{threeparttable}
		\caption{Simulation Environment Comparison}
		\label{tab:features}
		\begin{tabular}{@{}lc GG !{\kern1em} GGG !{\kern1em} GG@{}}
			\toprule
			Simulator  & Cost  & \multicolumn{18}{c}{Features \& Requirements} \\
			\midrule
			&&	\rot{Builtin Mobility}
			& 	\rot{Optical networking}
			& 	\rot{5G}
			& \rot{Full Stack}
			& \rot{Scriptable}
			& \rot{Packet capture}
			& \rot{Local simulation}
			& \rot{Cloud service}
			& \rot{GUI based}
			& \rot{Configurable Network Traffic}
			& \rot{Abnormal Conditions Injection}
			& \rot{Sumo Integration}
			& \rot{Packet Loss}
			& \rot{Traffic Grooming}
			& \rot{Latency}
			& \rot{Operating System}
			& \rot{Batch operations}
			& \rot{Unrestricted Element Count}\\
			\midrule
			\ex{Opensource}{     }  {  }          				                {444}{444} {444}{444}{444} {444}{444}\\
			\ex{\hspace{2mm} NS3 (GNS3)}{ }	   			{Free}               	{222}{333} {303}{332}{313} {022}{733}\\
			\ex{\hspace{2mm} NS2}{ }					{Free}                  {222}{333} {300}{332}{313} {022}{633}\\
			\ex{\hspace{2mm} OMNeT++}{ }				{Free}                  {222}{233} {302}{222}{313} {022}{733}\\
			\ex{\hspace{2mm} QualNet}{ }				{Free}                  {203}{133} {302}{222}{313} {022}{533}\\
			\ex{\hspace{2mm} ONS}{ }					{Free}                  {133}{113} {303}{222}{323} {022}{533}\\
			
			\midrule
			\ex{Vendor}{ }{ }       								  		    {444}{444} {444}{444}{444} {444}{444}\\
			\ex{\hspace{2mm} NETSIM}{\dag*}				{Yearly 4500+} 			{223}{213} {333}{202}{323} {002}{500}\\
			\ex{\hspace{2mm} ADVA}{ }					{Quota based}			{223}{212} {333}{212}{313} {022}{502}\\
			\ex{\hspace{2mm} Optiwave Systems}{\dag*}	{Yearly 3200+}          {221}{212} {333}{120}{133} {022}{502}\\
			\ex{\hspace{2mm} MATLAB (Simulink)}{\dag}	{Yearly 600-2400}       {222}{233} {303}{322}{022} {022}{702}\\
			\bottomrule
		\end{tabular}
		\begin{tablenotes}
			\item \hfil
			$\feature3=\text{Natively available}$; 
			$\feature2=\text{Indirectly available}$; 
			$\feature1=\text{Externally available}$;
			$\text{\feature0}=\text{Not available}$;
			\item \hfil
			$\text{W,U,B}=\text{windows,Ubuntu,Both}$;
			\textsuperscript{\dag}has academic discounts;
			\textsuperscript{*}end-user tool available
	\end{tablenotes}    \end{threeparttable}
	\label{comparisontable}
\end{table*}

Starting with the open-source simulators, NS3 and its windows variation GNS3 boast a large scientific following. This is because the system developed a robust base that attracted many users to build their dedicated simulation tools on top of it. In addition, the ongoing support of the community makes it an increasingly attractive tool for newcomers and veteran researchers. For 5G \cite{aqeeli2017power} and OTN-based simulations, there are several NS3/GNS3-based solutions available.

\begin{itemize}
	\item 5G-LENA is one of the NS3-based modules that can simulate 5G New Radio (NR) cellular networks. The simulator is an evolution of the famous LENA, the LTE/EPC Network Simulator. It is currently being developed and maintained by the Mobile Networks group CTTC (Centre Tecnològic de Telecomunicaciones de Catalunya). It is a full stack simulation of the 5G NR setup that allows the user full utility and control over the simulation depth and configurations.	
	\item Photonic WDM Network Simulator (PWNS) is one of the NS3-based modules used in the simulation of OTN-based networking. While the tools it offers are of great use, the lack of development and community updates of its base hindered its popularity as it is difficult to integrate it with more current NS3 releases.
\end{itemize}

OMNet++, similar to NS3, is a general networking simulation environment with a vital support and development community. However, unlike NS3 or GNS3, it boasts a more friendly and robust GUI-based interface, making it a more attractive option for development.
\begin{itemize}
	\item Simu5G OMNeT++ module allows researchers to simulate and benchmark solutions on an easy-to-use framework.  
	It offers support to optimization tools such as CPLEX and can be integrated with other modules from the INET Framework, allowing network scenarios where 4G and 5G coexist.
	\item The optical networking OMNeT++ module allows researchers to implement OTN-based topologies with unique structures, such as Optical switches, amplifiers, etc. In addition, the advantage of fully controlling the placement and capabilities of each component allows for better testing and integration with other solutions during testing. Moreover, its support community remains active with slight compatibility issues with the latest OMNeT++ releases.
\end{itemize}

On the other hand, for vendor-based simulators, NetSim is one of the industry’s leading paid 5G NR simulation tools. It offers End-to-End simulation of 5G networks, GUI-based with drag and drop capabilities, packet animator, results in dashboard, packet-level simulation with detailed packet trace, event trace, and NR log file generation. Additionally, it offers a fully controllable app-related user behavior and integration with mobility-based solutions such as SUMO. Unlike open source solutions, it allows for cloud-based subscription, letting researchers develop the topologies locally but run the simulation environments on the cloud, forgoing the need for advanced hardware requirements.

The vendor and open source features offer an attractive combination, but it becomes cumbersome for complete end-to-end simulations. For example, suppose the user opts for a vendor-based solution. In that case, they could be constrained by the number of users included in the license, limiting the size of their simulations, forcing multiple iterations to achieve desired results, or forced to purchase additional packages. On the other hand, if the user opts for an open-source solution, they will face setup-related issues and possible hardware-based limitations, causing similar simulation size-related challenges. 

\subsection{Contribution}
\indent\indent To address these issues, we propose a modular-based simulation system that maximizes the user’s ability to simulate 5G and OTN-based environments. The system is set up to allow multiple devices to run concurrently to achieve a more extensive encompassing simulation. Additionally, it allows the mixed use of multiple solutions, further increasing the user’s throughput and adaptability. This is achieved by exploiting the use of packet files as both input and output in addition to a controlling file-triggered script to monitor and trigger the various system modules when ready.

The remainder of this chapter is organized as follows: Section \ref{ch4_sim_description} describes the simulator including all of its internal modules. Then, Section \ref{ch4_performance} evaluates the performance of the proposed and developed simulator using multiple testing methods. Moreover, it discusses the achieved results. Finally, Section \ref{ch4_conc} concludes the chapter.

\section{Simulator Description}\label{ch4_sim_description}
\begin{figure*}[ht]
	\centerline{\includegraphics[scale=0.48]{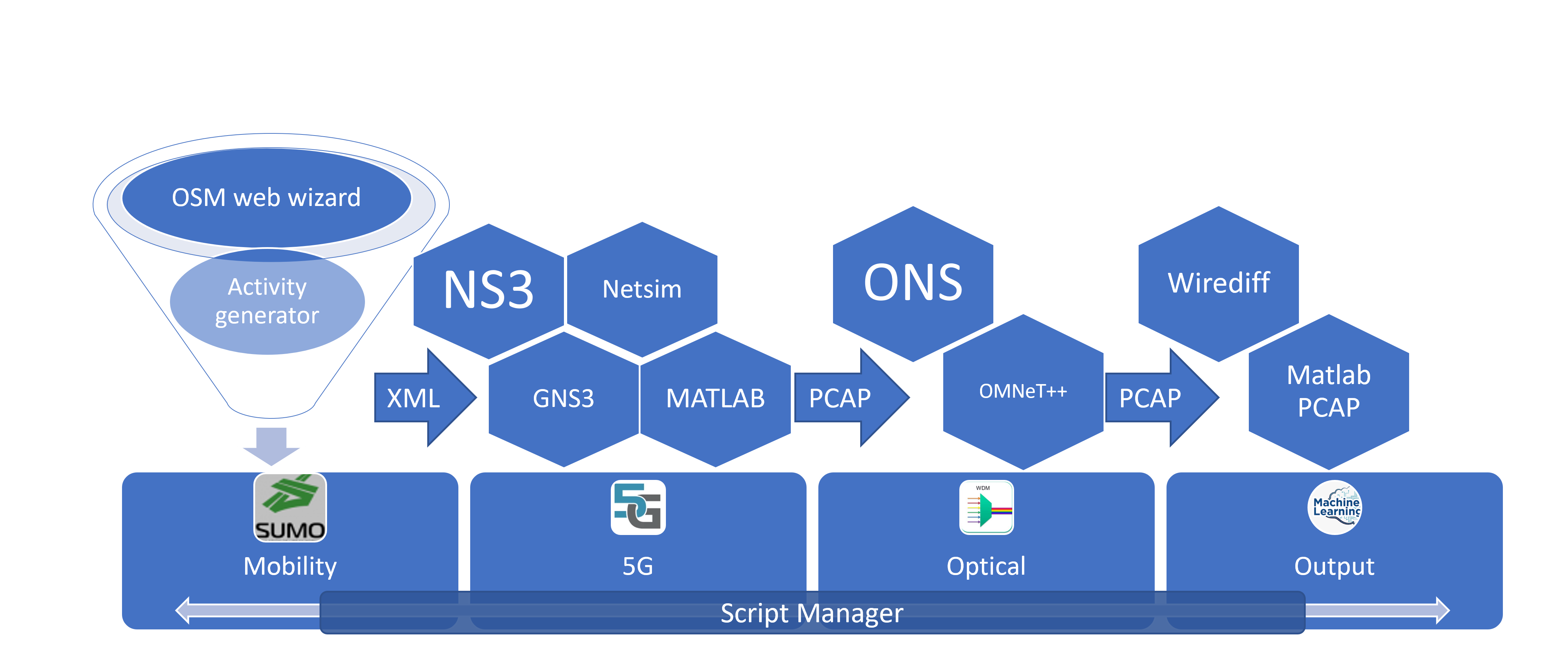}}
	\caption{System Design.}
	\label{fig}
\end{figure*}

\indent\indent The system comprises four independent components to maintain its modular nature, with the outputted files being the only interaction between them, as shown in Figure \ref{fig}. This allows the users to capitalize on the systems’ two main benefits. First, it allows for both single and distributed operations by relying on the script to direct and synchronize the number of devices used using file creation as triggers. Second, the system allows multiple simulation methods for 5G and OTN.
In what follows, a brief description of each of the four modules is provided.

\subsection{Mobility module}
\indent\indent This module is responsible for mimicking the real-life nature of the mobility used in the simulator. This is achieved by using SUMO and OSM wizard. The user begins by isolating the map section to be simulated using the OSM web interface and allocating the number of vehicles and pedestrians. Once generated, the user can further enhance the quality of the mobility simulation by editing the city properties XML files to replace generic values and assign more realistic ones. This creates sumo-compatible demands for a synthetic population by adjusting population-related aspects.
\begin{itemize}
	\item Population:
	\begin{itemize}
		\item Inhabitants and demographic division
		\item Households distribution
		\item Employment patterns
	\end{itemize}
	\item Location-based:
	\begin{itemize}
		\item Bus lines and stops.
		\item Schools, workplaces, malls.
	\end{itemize}
	\item Time-based:
	\begin{itemize}
		\item Work hours 
		\item Education hours
		\item Opening/ losing times for specific locations like Malls, stadiums etc. 
	\end{itemize}
\end{itemize}

This can also help generate abnormal test events as the user can create massive congestion or everyday public events. This is important for proper testing of ML-based solutions to ensure the training and testing include aberrant events improving the quality of the resulting models.
\subsection{5G module}
\indent\indent This module attaches 5G-based traffic demands to the mobility captured earlier. Unlike the mobility generation stage, the 5G module can leverage pipelining techniques to increase the simulation output volume. This is done in multiple ways based on the 5G simulator used. The most common open-source application for this is NS3 or its windows-based variation GNS3, and the vendor-based Netsim application. The script first prompts the user to check for the mode of operation, either singular or distributed, and choose the available applications. Based on that selection, if the language is fully scriptable (as outlined in Table \ref{comparisontable}), the script can invoke the application using the sumo file as bases or, in the case of Netsim, can prepare and guide the user through the GUI-based steps necessary to get the simulation started. 
In either mode, the PCAP file generation will be capped using the XML sumo file outlining the number of communication elements in each setup. This will trigger subsequent steps once that number of PCAP files is achieved or a timeout is reached.

Once the simulation is finalized, the PCAP files are collected and renamed using batch renaming software to conform to each app’s varying naming convention in cases where the file naming is not fully customizable. The output is then saved in the master simulation folder.

\subsection{OTN module}
\indent\indent Currently, the OTN module has been tested only on OMNeT++. The script feeds the PCAP files using some of the software’s built-in functions. The OTN network is built using disjointed pairs to allow various path variations to occur concurrently. The OMNeT++ simulator offers a fully customizable OTN structure, including editable components such as optical amplifiers, switches, lines, receivers, etc. The traffic grooming is generic but can be customized using community-based libraries accessible to the users. Once the resulting PCAP file reaches the final destination node, the script captures the file edit and triggers the OTN output. 

Each PCAP file stream is truncated to isolate only the initial input and final output. It is then renamed based on the User ID initiating node and destination nodes. This information is used later on and extracted by the script to generate the networking latency metrics.

\subsection{Output processing}
\indent\indent After the PCAP files begin populating from each simulation environment, the script can trigger idling local resources to start processing and extract relevant metrics such as latencies, packet loss, error rates, etc. This is done using two methods: MATLAB for scripting only, and Wirediff for more graphic and open-source options. Once the data has been collected, it is saved as either a MAT File or a TXT file based on the user selection.

\section{Performance evaluation}\label{ch4_performance}
\subsection{Objective}
\indent\indent To ensure the system’s usability for ML purposes, we need to ensure that its overall capabilities match or exceed those of its individual components working separately. To best test, two main attributes are identified as the focus of evaluating the quality of the data collected as a whole and the number of differences between each variation. This is done to ensure that the data the variations output is cross-compatible, a crucial attribute to ensure the correctness of any ML-based solutions built based on the dataset.
\subsection{Metrics}
\indent\indent The metrics used to measure the effectiveness of the simulator are chosen to offer a comprehensive view of the system’s functionality. The three metrics are: 
\begin{enumerate}
	\item The similarity of the data outputs of the different 5G module variations in the form of an encompassing conformity score.
	\item The simulation time required for the components tested for both monitored and automated portions to highlight the human factor required.
	\item The amount of data collected during those simulations in the form of users per simulation session of comparable length.
\end{enumerate}  

\subsection{Testing methods}
\indent\indent To properly test the system capabilities given the intended accessibility, the testing is done on three desktop devices with identical capabilities limited to average specifications. The testing is done for both singular and distributed setups. The variations tested are limited to the windows-based solutions with GNS3, and Netsim used for the 5G traffic simulation, while the OTN is handled using OMNeT++.
The following tests are conducted:
\begin{itemize}
	\item Singular mode with a static sumo environment running independently on multiple 5G-only environments running a unified single application type on all users. 
	\item Singular mode with a static sumo environment running independently on multiple variations of the entire system. 
	\item Distributed mode with a static sumo environment running collaboratively on both the cloud Netsim and NS3 variations.
	\item Distributed mode with multiple sumo environments running collaboratively on the cloud Netsim and NS3 variations.
	\item Distributed mode with multiple sumo environments running on NS3, running collaboratively.
\end{itemize}

The above combination of tests provide full coverage of the most common use scenarios of the system with a focus on the volume of data produced, its conformity with variations, and the quality it provides for use in ML discussed below.
\subsection{Results}
\begin{figure}[!ht]
	\centerline{\includegraphics[scale=0.95]{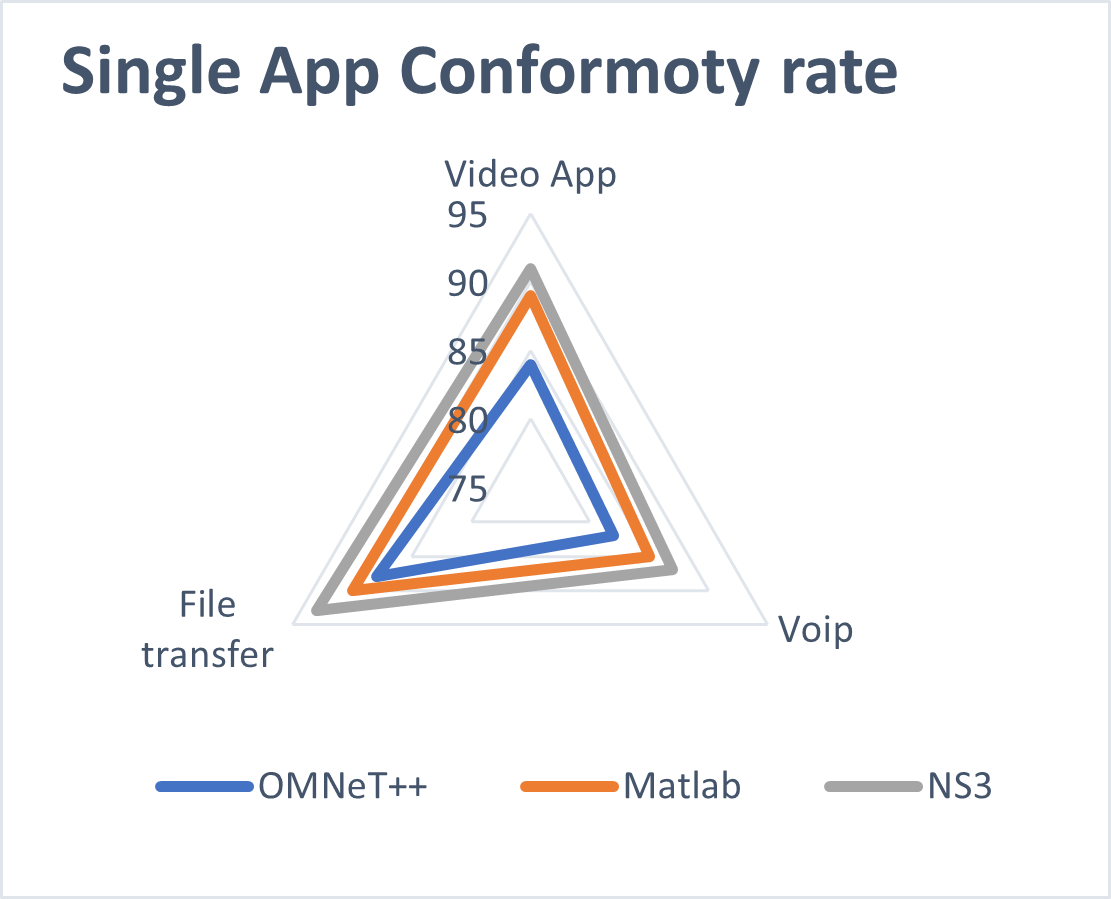}}
	\caption{Single app conformity compared to Netsim.}
	\label{fig2}
\end{figure}

\indent\indent To showcase the system viability, we first test the output of the 5G generation module variations to ensure their output can exist within the same data stream for ML usage without creating any biases or related issues. Figure \ref{fig2} shows the output conformity measured using  equation \ref{conform_eqn} to best represent the output inter compatibility of the various applications such as simulating Video streaming, VOIP, and File transfer services on all users in the following environments: NS3, OMNeT++, and NetSim. The impact of conformity has been highlighted in a number of machine learning-focused research efforts such as \cite{yang2021lightweight}, where the challenge of concept drift was discussed and the impact the quality of datasets used has on the model accuracy. The results show high conformity of the data outputted across all three variations compared to Netsim, with a slight reduction in VOIP traffic. These high levels are achieved due to the highly configurable nature of the application deployment in all three applications, with NS3 and OMNeT++ offering the most control.

\begin{equation}\label{conform_eqn}
\begin{split}
Conformity  =  6(Latency_{req}) + 4(Demand_{dur}) \\ + 2(Demand_{freq})+ 3(Packet_{Avg Size})
\end{split}
\end{equation}

\begin{figure}[!ht]
	\centerline{\includegraphics[scale=0.95]{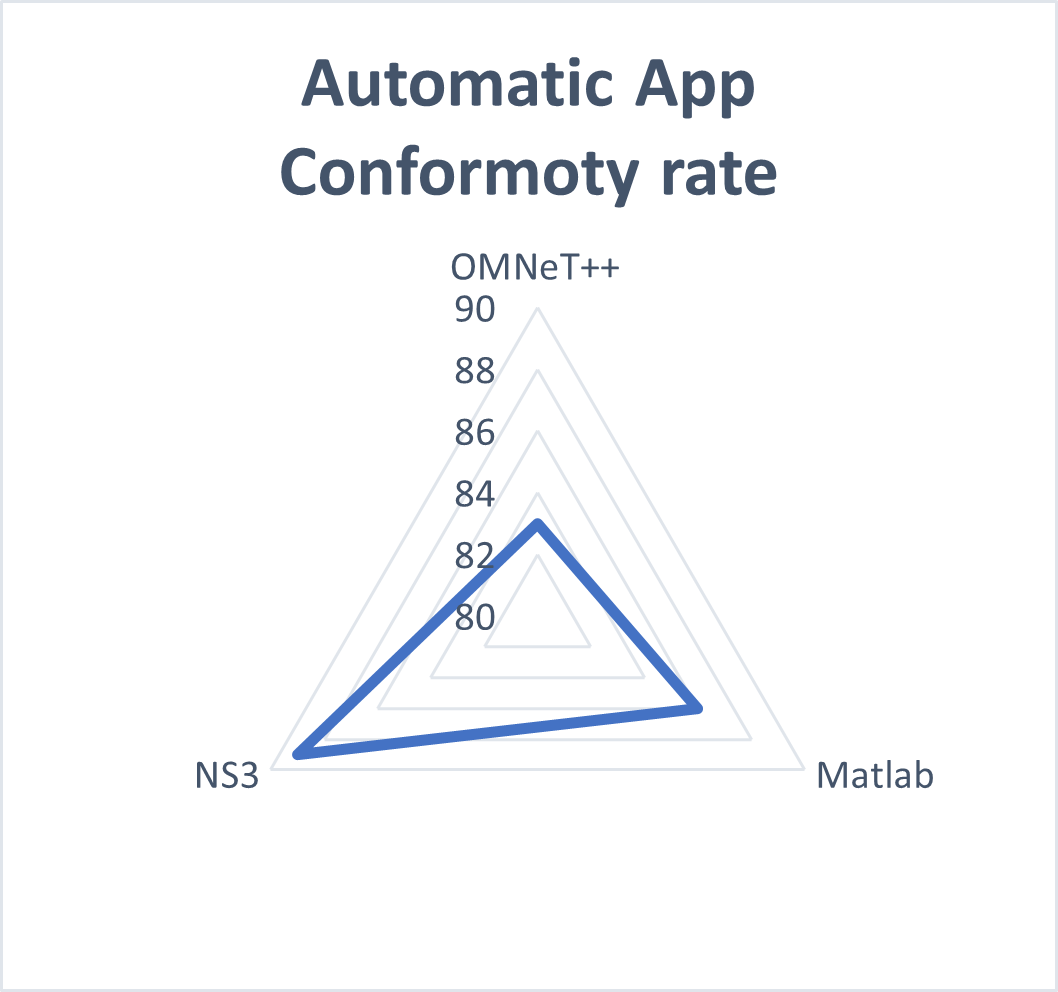}}
	\caption{Heterogeneous app conformity compared to Netsim.}
	\label{fig3}
\end{figure}

However, manually adjusting the application deployment is neither scalable nor compatible with the system's goal of low maintenance. Therefore, to test the system’s automatic attachment of applications to users based on the demographic data supplied in SUMO, we test the system mentioned above. Figure \ref{fig3} shows the system’s conformity after allowing them to use built-in functions applicable to manage the application distribution among the users. Results show slightly lower conformity but remain within an acceptable range for ML purposes. The increased variation can be positive, especially when building resilient ML-based systems.
\begin{table}[b]
	\footnotesize
	\caption{5G Simulator Output Volume}
	\centering
	\begin{tabular}{l|c|}
		\cline{2-2}
		& \multicolumn{1}{l|}{\textbf{Number of Users}} \\ \hline
		\multicolumn{1}{|l|}{\textbf{OMNeT++}} & +900                                          \\ \hline
		\multicolumn{1}{|l|}{\textbf{Matlab}}  & +700                                          \\ \hline
		\multicolumn{1}{|l|}{\textbf{NS3}}     & +900                                          \\ \hline
		\multicolumn{1}{|l|}{\textbf{Netsim}}  & 500                                           \\ \hline
	\end{tabular}
	\label{table2}
\end{table}

After ensuring that multiple system variations are viable, we shift our focus towards the quantity of data generated. Two main cases are in consideration: 
\begin{enumerate}
	\item Creating multiple runs on a single geographic area best suited for testing related to 5G and edge computing.
	\item Creating a massive simulation over multiple geographic areas.
\end{enumerate} 
 
\begin{figure}[ht]
	\centering
	\includegraphics[scale=0.95]{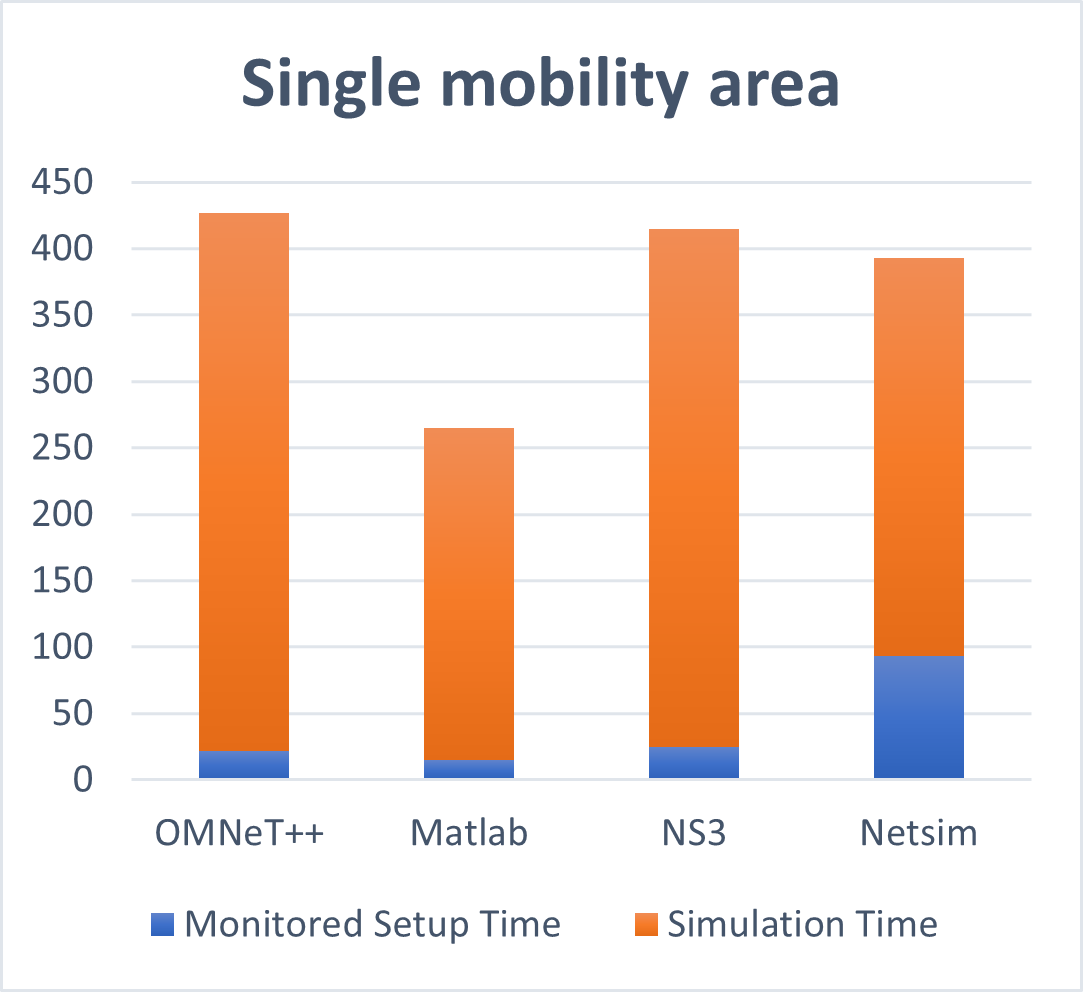}
	\caption{Simulating iteration of a single limited size environment.}
	\label{fig4}
\end{figure}

\begin{figure}[ht]
	\centerline{\includegraphics[scale=0.95]{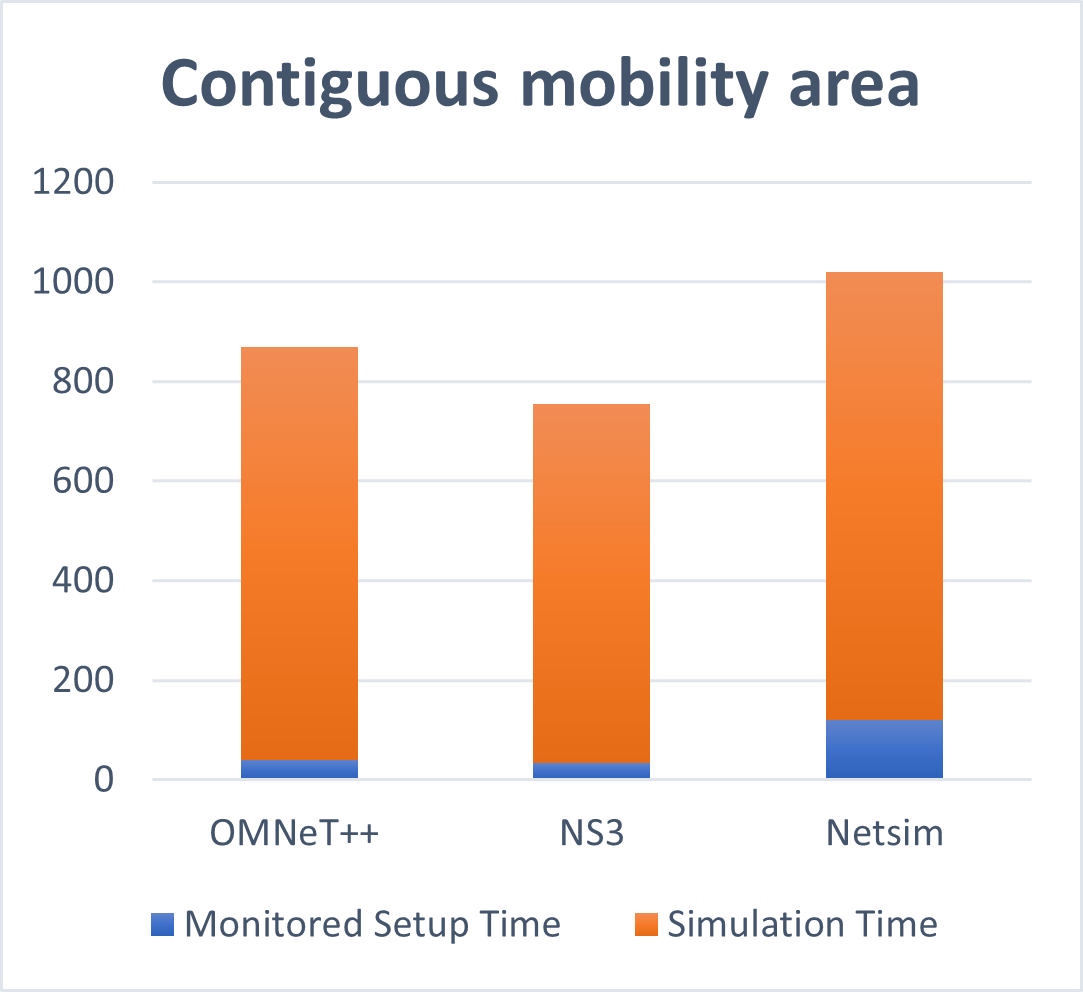}}
	\caption{Simulating a single large environment.}
	\label{fig5}
\end{figure}

Figures \ref{fig4} and \ref{fig5} compare the aggregate simulation runtimes, including the duration of required user interaction for hybrid and pure open-source distributed-based solutions. The results illustrated in Figure \ref{fig4} show that the use of Netsim required the most monitored setup time. At the same time, it offloaded the need for local hardware resources but required more user intervention in the setup stage due to its method of sumo mobility extraction. On the other hand, Figure \ref{fig5} shows the NS3-based solution, and its highly scriptable nature allowed for better-automated simulation. Another aspect to consider, shown in Table \ref{table2}, is the main shortcoming of using Netsim due to licensing-related limited number of simulated users. While it is considered a hindrance to the local solution and offers a slight increase in the number of simulated users, it requires more time and reservation of the resources.

\section{Conclusion}\label{ch4_conc}
\indent\indent In this chapter, we presented a viable simulation system that uses a combination of state-of-the-art environments to address the ongoing issue bogging down the development of AI-based systems in next-generation networking environments, especially in 5G and optical networking. The system considered combinations of paid and open-source solutions to work seamlessly to achieve the highest diversity and volume of data possible. The system testing results show considerable improvements over traditional approaches with minimal increase in the requirements or the need for dedicated hardware. The system can be improved further by developing the pipelining apparatus controlled by the script to allow for better handling of the live feed of PCAP files while the simulation’s earlier stages are ongoing to reduce the needed runtimes further. In addition, packaging the system using OS or VM images reduces the necessary setup and know-how to create customized datasets.
\bibliographystyle{IEEEtran}
\bibliography{main}
\end{document}